\documentstyle[aaspp4]{article}
\newcommand{\eg}{e.g.,~}
\newcommand{\ie}{i.e.,~}
\newcommand{\etal}{et al.~}
\newcommand{\um}{\mbox{$\,\mu\rm m$}}

\begin{document}

\title{Mid-Infrared Emission Features in the ISM: \\
	  Feature-to-Feature Flux Ratios}
\author{Nanyao Y. Lu}
\affil{Infrared Processing and Analysis Center\\
        MS 100-22, Caltech\\
        Pasadena, CA 91125\\
        Email: lu@ipac.caltech.edu}

\begin{abstract}
\indent
Using a limited, but representative sample of sources in the ISM 
of our Galaxy with published spectra from the {\it Infrared Space 
Observatory}, we analyze flux ratios between the major mid-IR 
emission features (EFs) centered around 6.2, 7.7, 8.6 and 11.3\um,
respectively.  In a flux ratio-to-flux ratio plot of 
EF(6.2\um)/EF(7.7\um) as a function of EF(11.3\um)/EF(7.7\um),
the sample sources form roughly a $\Lambda$-shaped locus which
appear to trace, on an overall basis, the hardness of a local 
heating radiation field.  But some driving parameters other than
the radiation field may also be required for a full interpretation
of this trend.  On the other hand, the flux ratio of 
EF(8.6\um)/EF(7.7\um) shows little variation over the sample sources,
except for two HII regions which have much higher values for this 
ratio due to an ``EF(8.6\um) anomaly,'' a phenomenon clearly 
associated with environments of an intense far-UV radiation field.
If further confirmed on a larger database, these trends should 
provide crucial information on how the EF carriers collectively 
respond to a changing environment.  

\end{abstract}


\section{Introduction} \label{sec1}

Since they were first discovered about two decades ago (Gillett 
\etal 1973), the mid-IR emission features (EFs), \ie those broad 
emission bands centered respectively at 6.2, 7.7, 8.6 and 
11.3\um,  have been detected from a variety of sources in our 
own Galaxy as well as in some other galaxies.   This makes it 
rather clear that the carriers
of these EFs play a significant role in regulating the physical 
conditions in the ISM.  One concept that has gained popularity 
is that these EFs arise from the vibrational modes of the so-called
aromatic hydrocarbon molecules (hereafter PAHs;  L\'eger \& Puget
1984; Allamandola \etal 1985).  However, this is largely based
on a wavelength coincidence between the observed EFs and 
the absorption bands measured in laboratories, a criterion 
that other candidate EF carriers (\eg Papoular \etal 1989; 
Sakata \etal 1984) also seem to satisfy.   It is therefore
important to test this PAH and other candidate scenarios in as 
many ways as possible.  One possible way is to observe how the strength 
ratios between the EFs react to a changing heating radiation field
in the ISM.  This approach is viable because, for example, if EFs
arise from PAHs, some of their feature-to-feature strength ratios
should depend on the hardness of the heating radiation field via
factors such as the fraction of PAH cations (\ie singly ionized 
PAHs;  Langhoff 1996) and the degree of dehydrogenation (\eg Jourdain
de Muizon \etal 1990).  Because of the limited sensitivity and 
spectral coverage associated with sub-orbital platform observations,
studies on how features respond to a changing environment have 
been carried out so far only to a limited extent (\eg Cohen \etal
1986; Joblin \etal 1996).

With its unprecedented sensitivity and continuous spectral coverage,
the Infrared Space Observatory (ISO; Kessler \etal
1996) has been used to obtain mid-IR spectra of sources in a 
variety of environments ranging over far-UV intense to typical
diffuse ISM.   This makes it possible for the first time to 
study the relative strengths among all the major EFs under a 
wide range of physical conditions.  We gather here the published 
ISO mid-IR spectra on a limited, but representative set of sources
in the ISM of our Galaxy to seek possible correlations between 
the feature-to-feature flux ratios and the local heating radiation
field.  While the former parameters are measured directly from 
the ISO spectra, the latter is inferred from the properties of 
the stars that likely dominate the radiation field at 
the locations where the ISO spectra were taken.

In Sect.~2 we describe the sample, the data and how we evaluate
the integrated flux of a feature.  In Sect.~3 we analyze 
feature-to-feature flux ratios and identify some possible systematic
trends.  Some implications from these trends are discussed in 
Sect.~4, when applicable, in comparison with the current knowledge
of PAHs.

\section{The Data} \label{sec2}

Our sample sources are listed in column~(1) of Table~1, grouped
into the following categories based on how hard their local 
radiation fields are likely to be:
(1) the five compact HII regions with the highest S/N ratios
in Roelfsema \etal (1996) and the HII region in the M17 complex
from Verstraete \etal (1996); (2) two photodissociation regions 
(PDRs) in the M17 region also from Verstraete \etal (1996); (3)
three reflection nebulae: a cloud edge in Ophiuchus from Boulanger
\etal (1996), NGC$\,$7023 from Cesarsky \etal (1996a), and 
vdB$\,$133 from Uchida \etal (1998); and (4) the two brightest
spectra taken along the Galactic plane in Mattila \etal (1996).
To supplement this last category of the weakest sources in the 
sample, we also included two published spectra along the inner 
Galactic plane taken with IRTS from Onaka \etal (1996) and an ISO
spectrum from Boulade (1996) of the interstellar medium in the 
central part of NGC$\,$5195, a galaxy with very few stars younger
than B5.    The corresponding instrument is given in column~(2)
in the table, where SWS stands for the ISO short-wavelength spectrometer
(de Graauw \etal 1996), CAM-CVF refers to the ISO-CAM with its 
circular variable filters (Cesarsky \etal 1996b), and PHT-S refers
to the spectroscopic mode of ISOPHOT (Lemke \etal 1996).  Most of 
these ISO spectra were published as part of the first ISO results.
While their absolute flux scale and sky subtraction may need 
further improvement, the quoted relative accuracy over the wavelength
range of 6 to 12\um\ is better than 20\%.  To avoid the problem 
of spectral resolution inhomogeneity over the sample sources, we 
consider only the integrated feature fluxes that are drawn directly
from the published figures.

We subtract a linear ``continuum'' (in $F_{\lambda}$)
from each emission feature as follows:  For the emission feature
at 6.2\um\ [hereafter EF(6.2), similarly for the other features], 
this is a line connecting the mean spectral value around 5.9\um\ 
to that around 6.9\um; for EF(7.7) and EF(8.6), a line drawn in a 
similar way from 6.9\um\ to 9.9\um; and for EF(11.3), a line from 
10\um\ to 12\um.  Clearly, these continuum definitions are largely
subjective and somewhat oversimplified.  But a more accurate 
continuum definition should not change our results in a significant
way.  Also, for the PHT-S and IRTS spectra which extend only up
to about 11.6\um\ in wavelength, we determine only lower and upper
limits on the continuum for EF(11.3) as described in the next paragraph.

After the continuum subtraction, the spectrum of each EF is
binned into a histogram with a bin size of $\sim 0.1\mu$m\ in
order to evaluate its integrated flux.  The integration is 
done as follows:  the flux of EF(6.2) is an integration of 
the spectrum from 6.0 to 6.6\um\ in wavelength;  for EF(7.7),
from 7.2 to 8.1\um; for EF(8.6), from 8.1 to 9.9\um; and for
EF(11.6), from 10.5 to 12\um.   For the PHT-S and IRTS data, 
we determine only an upper limit and a lower limit on the flux
of EF(11.3) as follows:  The upper limit is measured from 
the spectrum after subtracting out a continuum with a fixed 
$F_{\lambda}$ equal to that around 10\um.  In this case, 
the part of EF(11.3) that is beyond the wavelength cutoff of
the spectrometer is accounted for by assuming a symmetric 
profile with respect to the wavelength of the feature peak. 
For the PHT-S data, the lower flux limit is obtained by integrating 
the spectrum up to 11.6\um\ after subtracting out a linear continuum 
in $F_{\lambda}$ connecting the mean spectral value around
10\um\ to that at 11.6\um.  For the low-resolution IRTS data
that appear to lose a large fraction of EF(11.3), we take a
loose lower flux limit that equals 60\% of the upper limit. 
The procedure used here should introduce an uncertainty of $\lesssim
20\%$ in the integrated flux.  Also, any systematic effect
of this uncertainty on the feature-to-feature ratios should 
be more or less uniform over the whole sample.

Our results are given in columns~(3) to (5) in Table~1 in terms
of the following feature-to-feature flux ratios: EF(6.2)/EF(7.7),
EF(8.6)/EF(7.7) and EF(11.3)/EF(7.7), where and hereafter in 
this letter we simply use, for example, EF(6.2)/EF(7.7) to refer
to the flux ratio of EF(6.2) to EF(7.7).  EF(7.7) is chosen to 
be the common denominator for minimizing, on an overall basis,
the wavelength baselines used in these ratios.  Also listed in 
Table~1 are the known or estimated spectral types of the dominant
heating stars in column~(6) and estimated mean dust temperatures
in column~(7).  For each of the compact HII regions, this spectral
type was derived, under a single-star heating assumption, by 
comparing the models of zero-age main sequence stars in Panadia
(1973) with a Lyman-continuum luminosity inferred from the total
IR luminosity of the HII region given in Table~1 of Roelfsema 
\etal (1996).  The slope of this linear Lyman-continuum/IR 
relation was determined using the measured number of the 
Lyman-continuum photons in the case of IRAS 18116-1646 (Garay 
\etal 1993).  The dust temperature is derived from a $\lambda^{-1}$
emissivity and the following far-infrared continuum fluxes: for
the compact HII regions and NGC$\,$5195, we used the 60 and 
100\um\ flux densities in the IRAS Point Source Catalog; for 
the 3 regions in M17, we used the 50 and 100\um\ maps in Gatley
\etal (1976); for NGC$\,$7023, we used the far-IR maps in Whitcomb
\etal (1981); for Ophiuchus and the spectra along the Galactic 
plane, we measured their 60 and 100\um\ surface brightness values 
on the IRAS Sky Survey Atlas plates; and for vdB$\,$133, the IRAS 
surface brightness values in Sellgren (1990) were used here.  
We emphasize that both the spectral types and dust temperatures 
given in Table~1 are for {\it indicative} purpose in this paper. 
In some cases, there is some uncertainty as to whether they 
represent the actual physical condition within the area where 
the mid-IR spectrum was observed.

\section{Feature-to-Feature Flux Ratios} \label{sec3}

Using these feature-to-feature flux ratios, we construct two 
pairwise plots in Fig.~1 where different symbols are used to 
differentiate the various source categories in Table~1:  
the 6 HII regions are represented by filled squares; the 2 PDRs
by open squares;  the three reflection nebulae by crosses; 
the 4 spectra along the Galactic plane by horizontal bars each
extending from the lower limit to the upper limit on the value
of EF(11.3)/EF(7.7) as given in Table~1; and the galaxy 
NGC$\,$5195 by an asterisk.  The typical errors are on 
the order of 30\% or less along either axis.  
There are two HII regions which are further circled in Fig.~1.
These are what we define below as sources with an ``EF(8.6) 
anomaly.''

\subsection{Relative Strengths between EF(7.7) and EF(8.6)}
\label{sec3.1}

Fig.~1a is a plot of EF(8.6)/EF(7.7) as a function of EF(11.3)/EF(7.7).
Apparently, the sources are distributed
into two distinct groups on this plot.  One group consists of 
the compact HII region IRAS$\,$18434-0242 and the HII region 
in M17.  These two sources are characterized by EF(8.6)/EF(7.7) 
$\sim 1.5$, a value which is at least three times as large as 
that of any source in the other group made of the other sample 
sources.  For the latter group, the ratios of EF(8.6)/EF(7.7) 
do not change significantly, with a group mean of $0.40$ and a 
standard deviation of only $0.07$.  Since the majority of the 
sources (both Galactic and extragalactic) with published mid-IR 
spectra so far belong more or less to the latter group, we refer
to the phenomenon that EF(8.6)/EF(7.7) $> 1$ as an ``EF(8.6) 
anomaly'' in this letter.

Both of the EF(8.6)-anomaly sources have a mid-IR continuum that
rises steeply toward longer wavelengths, about 3 to 5 times
steeper than the continuum of any of the other HII regions in 
the sample (see Roelfsema \etal 1996; Verstraete \etal 1996) 
or 2 to 3 times steeper than that of the hottest knot in 
the Antennae galaxies (Vigroux \etal 1996) when the spectral 
steepness is measured in terms of the ratio of the flux density
at 12\um\ to that at $6$\um.   This strongly suggests a 
presence, within the SWS aperture, of a very hot continuum from
dust grains heated by an 
intense far-UV radiation field in these EF(8.6)-anomaly sources.
A closer look at the spectra of the EF(8.6)-anomaly sources 
shows that EF(8.6) is not only stronger than EF(7.7) in peak 
intensity but also extends to a much wider wavelength range 
than that in any ``EF(8.6)-normal'' source.

\subsection{Relative Strengths among EF(6.2), EF(7.7) and EF(11.3)} 
\label{sec3.2}

In Fig.~1b, a plot of EF(6.2)/EF(7.7) as a function of EF(11.3)/EF(7.7),
the data points apparently form a locus that roughly resembles 
a $\Lambda$-shaped curve centered at EF(11.3)/EF(7.7) $\sim 0.3$. 
What may be significant is that the various source categories 
seem to occupy different segments of the curve: all the HII regions
but one (IRAS$\,$22308+5812) lie near the end of the curve 
characterized by small values for both EF(6.2)/EF(7.7) and 
EF(11.3)/EF(7.7);  the 2 PDRs are on the same side of the curve
as the HII regions, but located somewhat closer to the top of 
the curve where the three reflection nebulae are located; and on
the other side of the curve are the spectra taken along the inner
Galactic plane and that of NGC$\,$5195, presumably dominated by 
the emission from the diffuse ISM.

If the sources in the diffuse ISM category are indeed excited mainly
by somewhat late type stars,  our data suggests that there exists 
a well defined area in Fig.~1b for sources heated by a radiation 
field of a certain range of hardness.   The only clear exception 
to this interpretation is IRAS$\,$22308+5812, the HII region that
is in close proximity to the three reflection nebulae in Fig.~1b. 
This source actually has a mid-IR spectrum with little indication 
of a hot dust continuum rising toward longer wavelengths.  This is 
more like the spectra of reflection nebulae than those of the other
HII regions in the sample.  In this sense, its location in Fig.~1b
may not be very surprising.

On the other hand, if the dust temperature given in column~(7) of
Table~1 reflects the mean intensity of a radiation field, it seems
quite clear that the radiation intensity can not be the primary
driving force on how a source will be located in Fig.~1b.   For 
example, the overall radiation intensity in NGC$\,$5195 is 
comparable to those in some of the compact HII regions and is
certainly stronger than that in the Ophiuchus nebula.  But NGC$\,$5195
is clearly located closer to the other diffuse sources in Fig.~1b.
Another example comes from NGC$\,$7023 and the Ophiuchus nebula,
which lie close to each other in Fig.~1b.  These two sources have
a similar stellar energy distribution, but are exposed to a very 
different radiation intensity.

It should also be pointed out that, although the trend with the
radiation field in Fig.~1b seems to be quite significant from 
one source category to another,  it is still unclear how good 
it is within each source category.  For example, it is reasonable
to assume that the two EF(8.6)-anomaly sources represent the 
hardest UV fields in the sample, but they are not at the extreme
end of the trend in Fig.~1b.  Another example is vdB$\,$133, 
a reflection nebula that may be excited mainly by a star of 
spectral type F5 (Uchida \etal 1998).  However, in terms of EFs,
vdB$\,$133 is quite similar to the other two nebulae excited by 
much hotter stars.  As a result of the small numbers of sources
involved and the still large measurement errors, the certainty 
about these details of the trend in Fig.~1b is not as high as 
that about its overall pattern.  But it seems fair to say that 
although the heating radiation field plays a crucial role in 
determining the relative strengths of EFs, there may be other
regulators as well.

\section{Discussion} \label{sec4}

The result of Fig.~1a that the ratios of EF(8.6)/EF(7.7) for 
EF(8.6)-normal sources stay nearly constant is already quite secure
at this point.  This suggests either (a) that there is roughly a 
fixed ratio between the strengths of these two EFs or (b) that 
EF(7.7) has a long wavelength tail/plateau that dominates the 
integrated flux of EF(8.6) as a result of our dividing the two EFs
at 8.1\um\ in evaluating their fluxes.  Under the current framework
of PAHs, however, the ratio for EF(8.6)/EF(7.7) is expected to 
decrease with an increasing degree of dehydrogenation associated 
with an increasing hardness of the heating radiation field (Jourdain
de Muizon \etal 1990). So the laboratory results on PAHs favor
(b).

On the other hand, additional and improved data are probably needed
for further confirmation on the details of the distribution pattern
in Fig.~1b.  There is a slight possibility that, once more data 
points are inserted in Fig.~1b, a simpler pattern [\eg EF(6.2)/EF(7.7)
increases somewhat as EF(11.3)/EF(7.7) decreases] could emerge. 
However, if this $\Lambda$-shaped distribution pattern is further
confirmed, it could provide new constraints on the identification
of the EF carriers. For example, the right side of the $\Lambda$-shaped 
curve in Fig.~1b may be consistent with a picture where one has 
a combination of rising PAH temperature and increasing ionization
effect (Langhoff \etal 1996) as one goes up along the curve.  
However, the current knowledge of PAHs does not seem to suggest a
turnaround in EF(6.2)/EF(7.7).  An increasing photodestruction 
effect does suggest a decreasing ratio of EF(6.2)/EF(7.7) as smaller
PAHs are easier to be destroyed than their larger cousins 
(Allamandola \etal 1989; Allain, Leach, \& Sedlmayr 1996a, 1996b)
and as larger PAHs reach lower peak temperatures after absorbing 
a UV photon (L\'eger \& Puget 1989). But the wavelength difference
between the two features is so small that the dependence of their
ratio on the PAH size distribution seems inadequate to explain 
the observed large change in EF(6.2)/EF(7.7).  Besides, this effect
has to be counter-balanced by the fact that in an increasingly 
harder UV-rich environment, the average PAH temperature also gets
hotter.  Perhaps, this implies that something in addition to the 
radiation field is needed to explain the distribution pattern in 
Fig.~1b.

The laboratory counterpart of the EF(8.6)-anomaly phenomenon is 
unknown at this point.  One speculation is that both EF(7.7) and
EF(8.6) are stronger relative to the other features in a far-UV
rich environment.  Perhaps, the two features sit on top of an emission
plateau whose contribution to the fluxes of EF(7.7) and EF(8.6)
becomes relatively important only under an intense far-UV radiation
field.  In fact, such an emission plateau may put the EF(8.6)-anomaly
phenomenon in a sequence between those EF(8.6)-normal sources and
those with even more ``unusual'' EF spectra, \eg an ISO spectrum
dominated by a previously unknown broad feature around 8\um\ seen
in an ultra compact HII region (Cesarsky \etal 1996c) to those 
featureless spectra seen in many active galactic nuclei (\eg Aitken
\& Roche 1985).

Regardless of how consistent they are with the physical properties 
of a specific class of candidates for the EF carriers, the trends 
in Fig.~1 should inevitably lead us to a more complete picture on
how the relative EF strengths change from one type of environment
to another. A direct application of this would be to help interpreting
the global EF spectra from galaxies.  For example, a preliminary 
analysis shows that in a plot such as Fig.~1b, most normal galaxies
scatter within a region close to that occupied by those data points
taken along the inner Galactic plane (Lu \etal 1998), suggesting
that the global EFs of a galaxy may be dominated by the emission 
from the diffuse ISM component. 

\acknowledgments

The author is grateful to the anonymous referee for a number of comments
and suggestions that helped improving both the presentation and 
interpretation of the data, to L. Allamandola and D. Hudgins for
interesting conversations on PAHs, and to G. Helou, S. Malhotra and
C. Xu for helpful comments.  This work was supported in part by Jet 
Propulsion Laboratory, California  Institute of Technology, under a 
contract with the National Aeronautics and Space Administration.
%
%

\begin{figure}
\plotone{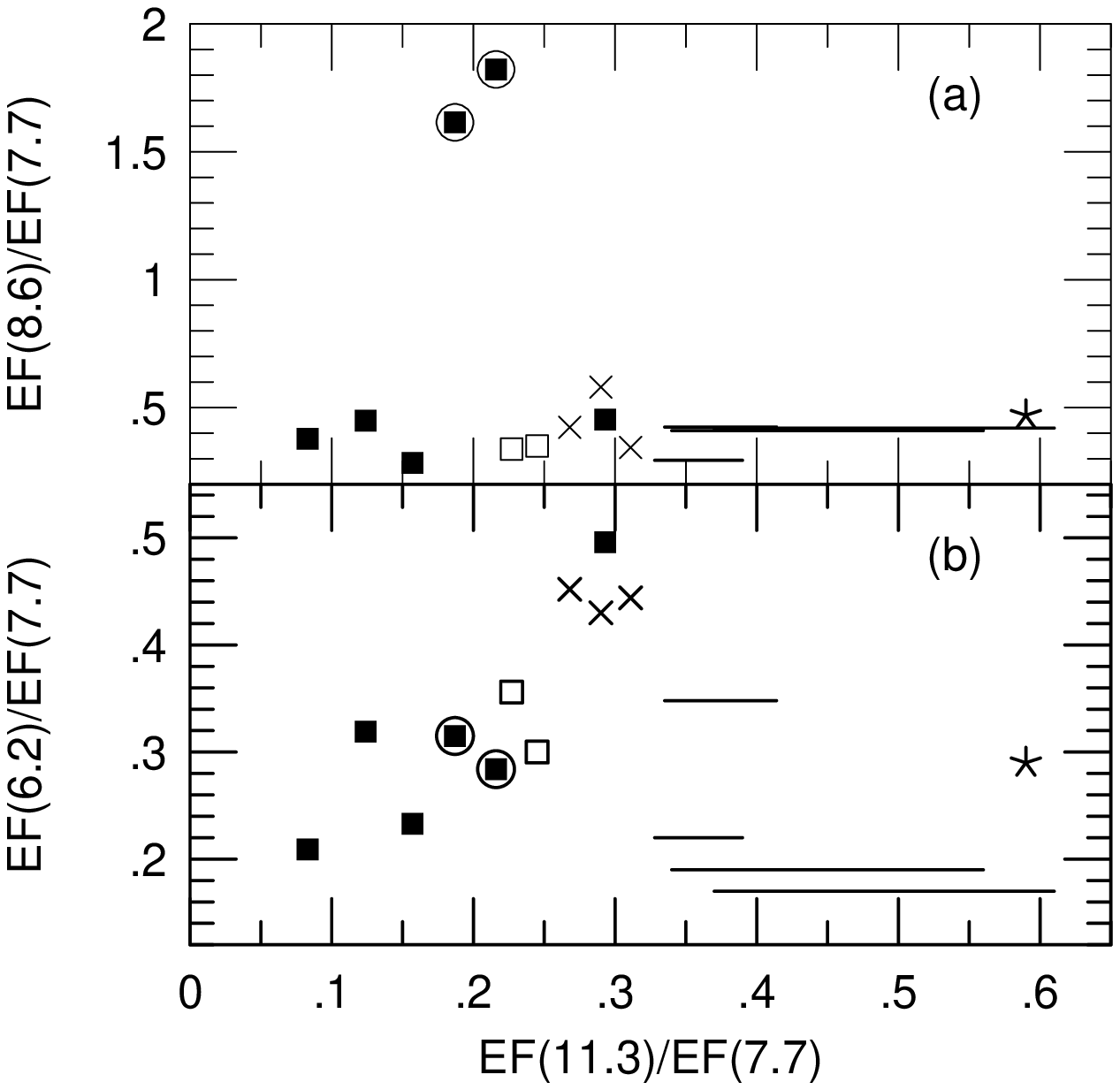}
\vspace{-2.5in}
\caption{
     Plots of feature-to-feature flux ratios for the sources listed in 
     Table~1: (a) EF(8.6)/EF(7.7) {\it vs.}~EF(11.3)/EF(7.7) and (b)
     EF(6.2)/EF(7.7) {\it vs.}~EF(11.3)/EF(7.7).  We use filled squares 
     to represent the 6 HII regions,
     with the two HII regions of the largest EF(8.7)/EF(7.7) ratios
     further circled; open 
     squares for the two PDR regions; crosses for the three reflection
     nebulae; horizontal bars for the diffuse ISM emissions
     along the Galactic plane, each extending between the lower and 
     upper limits on the value of EF(11.3)/EF(7.7) as given in Table~1;
     and an asterisk for the galaxy NGC$\,$5195.
     The typical errors are on the order of 30\% or less along either
     axis.}
\end{figure}
\newpage

\begin{deluxetable}{llccclc}
\scriptsize
\tablenum{1}
\tablewidth{0pt}
\tablecaption{Flux Ratios of Mid-IR Emission Features}
\tablehead{
\colhead{Source}  & \colhead{Instrument} & \colhead{(6.2)/(7.7)} & 
\colhead{(8.6)/(7.7)}  & \colhead{(11.3)/(7.7)}  & \colhead{Spec. Type} &
\colhead{$T_{dust}$} \\
\colhead{(1)}  & \colhead{(2)} & \colhead{(3)} & 
\colhead{(4)}  & \colhead{(5)} & \colhead{(6)} &
\colhead{(7)}}
\startdata
{\bf HII regions}: \nl
IRAS$\,$18434-0242$^a$    & SWS  & 0.32 & 1.62 & 0.19 & \ \ O6   & 38 \nl
IRAS$\,$18116-1646$^a$    & SWS  & 0.32 & 0.45 & 0.12 & \ \ O6.5 & 36 \nl
IRAS$\,$22308+5812$^a$    & SWS  & 0.50 & 0.45 & 0.29 & \ \ O7   & 38 \nl
IRAS$\,$19442+2427$^a$    & SWS  & 0.21 & 0.38 & 0.08 & \ \ O8   & 38 \nl
IRAS$\,$18162-2048$^a$    & SWS  & 0.23 & 0.28 & 0.16 & \ \ O9   & 40 \nl
M17 HII$^b$               & SWS  & 0.28 & 1.82 & 0.22 & \ \ O    & 58 \nl
{\bf PDRs}: \nl
M17 Interface$^b$\phn     & SWS  & 0.36 & 0.34 & 0.23 & \ \ O    & 48 \nl
M17 H$_2$-cloud$^b$       & SWS  & 0.30 & 0.35 & 0.24 & \ \ O    & 39 \nl
{\bf Reflect.~Nebulae}:\phn\nl
Ophiuchus$^c$\phn         & CAM-CVF      & 0.44 & 0.34 & 0.31 & \ \ B2 & 30 \nl
NGC$\,$7023 North$^d$\phn & CAM-CVF \phn & 0.45 & 0.42 & 0.27 & \ \ B3 & 50 \nl
vdB$\,$133$^e$		  & CAM-CVF      & 0.43 & 0.58 & 0.29 & \ \ F5/B7 & 30 \nl
{\bf Diffuse ISM}:\nl
($l=355^o$,$b=0^o$)$^f$        & PHT-S & 0.35  & 0.42 & 0.34$-$0.41  & \ \ ... & 27 \nl
($l=330^o$,$b=0^o$)$^f$        & PHT-S & 0.22  & 0.29 & 0.33$-$0.39  & \ \ ... & 27 \nl
($l\approx 44.3^o$,$b\approx -20'$)$^g$  & IRTS  & 0.17  & 0.42 & 0.37$-$0.61 & \ \ ... & 26 \nl
($l\approx 51.5^o$,$b\approx 1.5^o$)$^g$  & IRTS  & 0.19  & 0.41 & 0.34$-$0.56 & \ \ ... & 25 \nl
NGC$\,$5195$^h$		   	       & CAM-CVF  & 0.29  & 0.47 & 0.59 & \ \ ... & 35 \nl    
{\bf 
\tablenotetext{}{{\bf References}: (a) Roelfsema \etal (1996); (b) Verstraete \etal (1996);
		      (c) Boulanger \etal (1996); \newline
		  \null\hspace{0.8in} (d) Cesarsky \etal (1996a); (e) Uchida \etal (1998); 
		  (f) Mattila \etal (1996); \newline
		  \null\hspace{0.8in} (g) Onaka \etal (1996); and (h) Boulade \etal (1996).}}
\enddata
\end{deluxetable}

\end{document}